# On the relationship between entropy and information

Afshin Shafiee[*] and Majid Karimi
Department of Chemistry, Sharif University of Technology,
P.O.Box, 11365-9516, Tehran, Iran

*Abstract*

In this paper, we analyze the relationship between entropy and information in the context of the mixing process of two identical ideal gases. We will argue that entropy has a *special* information-based feature that is enfolded in the statistical entropy, but the second law does not include it directly. Therefore, in some given processes in thermodynamics where there is no matter and energy interaction between the system and environment, the state of the system may go towards a situation of lower probability to increase observer's information in environment. This is a kind of an information-based interaction in which the total entropy is not constrained by the second law.

## 1  Introduction

Gibbs' theorem and Maxwell's Demon are some fundamental and important topics which make us to be involved deeply in the concept of entropy [1, 2, 3]. In fact, in spite of the plenty of papers and books written in this area, the concept of entropy and especially its relation with information are still controversial subjects. Here, we are going to focus on this subject in a different way. We will review the well-known thought experiment of separating two ideal gases which is usually considered to explain Gibbs' theorem. Then, we will survey the partitioning process of an ideal gas to two equivalent parts. In the second experiment, although the process does not happen spontaneously, the entropy will increase for an isolated system. In fact, the latter experiment can be viewed as a reverse course of a relocation process for identical particles in which the entropy decreases because of the indistinguishability assumption suggested to resolve Gibbs' paradox. By an appropriate interpretation of this important process, we will propose a different generalization of the second law in material processes which is the main theme of this paper.

---

[*] Corresponding author: shafiee@sharif.edu

In section 2, we first review the content of Gibb's theorem. Then, in section 3, we will investigate Gibbs' paradox with a different point of view. The concept of the entropy and the second law of thermodynamic will be critically examined in section 4, where we propose two different notions of entropy, i.e., the thermodynamic entropy ($S_{th}$) and the statistical entropy ($S_{st}$). These two notions of entropy are discriminated when the material change happens. Subsequently, Maxwell's Demon will be analysed in our proposed framework in section 5 and we will sum up our results in the conclusion part.

## 2 Gibbs' theorem

Gibbs' theorem states that the entropy of an ideal gas mixture is equal to the sum of the entropies each pure gas would have, if it alone occupied the volume of the mixture at the temperature of the mixture [1]. Distinctly, for the mixture of two ideal gases of A and B, we have:

$$S_{A+B} = S_A + S_B \tag{1}$$

where, $S_{A+B}$ is the entropy of mixing and $S_A$ and $S_B$ are the entropies of two different ideal gases A and B. To discuss the accuracy of this theorem, the following thought experiment is usually suggested [4]. It is assumed that two ideal gases of A and B are mixed in a mobile cylinder in the volume *V* placed in a heat bath at the temperature *T*. At first, the whole state of the system constitute a single phase $\beta$ which is in mechanical and thermal equilibrium and is separated from the vacuum phase $\delta$ by an impermeable membrane (Fig.1).

There are also two other membranes, one permeable only to gas A and the other permeable only to gas B, in a way that we can gradually separate two gases by slowly pushing the cylinder to the right. If the number of particles of two gases are assumed to be the same at first ($N_A = N_B$), during the process two other phases $\alpha$ and $\gamma$ are formed which their partial pressures are related to each other as $P_A^\alpha = P_B^\gamma = \frac{1}{2}P_\beta$ and $P_A^\alpha + P_B^\gamma = P_\beta$, where $P_A^\alpha$ ($P_B^\gamma$) is the pressure of gas A (B) in phase $\alpha$ ($\beta$) and $P_\beta$ is the pressure of the mixture in phase $\beta$. This is the requirement of the mechanical equilibrium which we are assuming to be satisfied during the process. If the system is in mechanical equilibrium, the two phases $\alpha$ and $\gamma$ must have the same pressure and the sum of their pressures should be equal to the pressure of the intermediate phase $\beta$.

On the other hand, since the temperature is fixed, the change of the internal energy is zero ($\Delta U = 0$) and due to the assumption of mechanical equilibrium in both sides of the membrane, there will be no exchange of work with the environment (it is assumed that there exists no friction, so that a very little force is needed to move the membranes; i.e., $w = 0$). Also, the chemical potentials of the two gases will be the same in the adjacent phases, i.e.:

$$\mu_A^\alpha = \mu_A^\beta; \quad \mu_B^\beta = \mu_B^\gamma \tag{2}$$

In this situation, the system is in a state of material equilibrium too. Consequently, since the system is in total equilibrium during the whole process, each step will be reversible in its essence and hence $q_{rev} = 0$, where $q$ is the heat exchanged with the environment. Thus, $\Delta S = 0$ in the total process of mixing. This concludes the Gibbs theorem.

Gibbs' theorem has been always a subject of debate [5, 6, 7, 8]. However, we are not going to enter the details of such debates. The main point of our discussion here is that if the Gibbs thought experiment is used for two identical gases, it will result in a meaningful contrast between two notions of entropy we define as the thermodynamic entropy ($S_{th}$) and the statistical entropy ($S_{st}$). This, in turn, will clarify the relationship between the entropy and information.

Now consider the same argument as above for $2N_A$ identical particles of gas $A$ which constitute phase $\beta$ in a volume $V$ at temperature $T$ (Fig2). We still assume that system is in mechanical and heat equilibrium during the process. In an isothermal process, if one pushes the cylinder gradually to the right (assuming there is no friction again), two subsequent phases $\alpha$ and $\gamma$ each with $N_A$ particles of gas A in a volume of V and at temperature T will be formed. To do so, we assume that the membranes are sufficiently porous to permit the transition of particles A, but not generally so. This allows the gas to be slowly forced to move gradually from one part to another part. Essentially all the pressure changes occur in the porous partitions. Then, because the partitioning process is slow, pressure equilibrium is maintained at any intermediate state. Therefore, we have $P_A^\alpha = P_B^\gamma = \frac{1}{2}P_\beta$ and $P_A^\alpha + P_B^\gamma = P_\beta$ which the second relation approves the assumption of the mechanical equilibrium at both sides of the membranes. Also, we have $\mu_A^\alpha = \mu_A^\gamma$. However, it is clear that the system cannot be in a state of material equilibrium during the process. The transfer of matter from one part to another part is occurred irreversibly (and so the change of state is irreversible), since $\mu_A^\alpha \neq \mu_A^\beta$ and $\mu_A^\gamma \neq \mu_A^\beta$. Here, we have

$$\mu_A^\alpha = \mu_A^\gamma = \mu_A^0(T) + N_A kT \ln {P_A^\alpha}/{P^0} \qquad (3)$$

While on the contrary, $\mu_A^\beta = \mu_A^0(T) + N_A kT \ln {P_A^\beta}/{P^0}$, where $P^0$ is the standard pressure, k is Boltzmann constant and $\mu_A^0(T)$ denotes the chemical potential of gas A in a standard state.

On the other hand, in this process the temperature is constant (hence, $\Delta U = 0$) and the system performs no work on environment. Consequently, the process is adiabatic. Assuming the Boltzmann statistics for an ideal gas and considering the indistinguishability of the particles, from a statistical point of view, one concludes that:

$$S(\lambda N, V, T) = \lambda S(N, V, T) - \lambda k \ln \lambda^N \qquad (4)$$

where $\lambda$ is an arbitrary parameter. Thus, in this process, we have:

$$\Delta S = 2S(N_A, V, T) - S(2N_A, V, T) = 2k \ln 2^{N_A} > 0 \qquad (5)$$

i.e., the entropy change must be positive according to the statistical calculations. The second law states that for a closed system undergoing an irreversible adiabatic process, if $\Delta S > 0$, the process will happen *spontaneously*. The above process, however, shows a different situation, since the process does not happen spontaneously.

So, *why* should entropy increase, when the process is neither spontaneous nor there is any form of the exchange of energy with environment? This is a deep question which we are going to explore its dimensions in the following. Before that, however, we review Gibbs' paradox at a glance in the following section.

## 3  Gibbs' paradox

Considering the indistinguishability of particles, Gibbs showed that the entropy change in the process of mixing two identical ideal gases is zero [2]. To explain why entropy does not increase in this process, we place $N_A$ particles of gas *A* in each of the two separated cylinders having the same volume *V*. The two identical gases have no energy and material exchange with each other, but the overall system is placed in a heat bath at temperature *T*. Then, we isothermally expand each gas separately to volume 2*V*. Subsequently, assuming that there is no friction and no work exchange with the environment, we can mix the two gases by applying a negligible external force in a reverse fashion described before for the partitioning process in Fig. 2. The second step is a relocation process. In the first process (labeled as *a*) which is an isothermal expansion, $\Delta S_a = 2N_A k \ln 2$. According to the relations (4) and (5), for the second process (labeled as *b*) one gets:

$$\Delta S_b = S(2N_A, 2V, T) - 2S(N_A, 2V, T) = -2k \ln 2^{N_A} \qquad (6)$$

Therefore, $\Delta S_a + \Delta S_b = 0$. For canceling out the sum of the entropy changes in mixing process of two identical ideal gases, the assumption of no heat and work exchange with environment in process *b* is obligatory. This is the way that Gibbs' hypothesis of the indistinguishability of identical particles is shown to be relevant in resolving the paradox. However, the interesting point in the second process is that the entropy decreases in an adiabatic process. We stress again that this step is the reverse process of partitioning an ideal gas demonstrated in Fig. 2 (when one only replaces *V* with *2V*) and should not be confused with a compression process.

Now, what really happens in process b is that one interblends two different parts (occupied by the same number of particles) of a given ideal gas, combining them in one part only, without any energy cost. The combined final state, however, is a less probable state (compared to the initial state where the particles see the whole space of container) in which the observer gains more information about the location of particles. Hence, whereas the system is remained in an isolated status during the second process (i.e., no exchange of energy is involved there), the observer's information is changed.

Accordingly, we reach the same previous question (but in a different way) as to whether the entropy decrease in such a process (or increase in the reverse process) imply the violation of the second law of thermodynamic. Our answer to this question is negative. Nevertheless, to explain why our answer is consistent and coherent, we must provide a different meaning for the concept of entropy and the application of the second law of thermodynamic.

## 4  Entropy and the second law of thermodynamic

In a more general form of process *b*, if one mixes $\lambda$ identical ideal gases of type A to change the state of the system from $(N_A, V, T)$ to $(\lambda N_A, V, T)$, regarding the relation (4), one will obtain the entropy change as:

$$\Delta S_{IGM} = -\lambda k \ln \lambda^{N_A} \tag{7}$$

where *IGM* stands for "Identical Gas Mixing". The relation can be written as $\Delta S_{IGM} = k \ln \left(\frac{1}{\lambda}\right)^{\lambda N_A}$ where the term $\left(\frac{1}{\lambda}\right)^{\lambda N_A}$ is the probability that $\lambda N_A$ identical particles of *A* are being in one of the $\lambda$ possible states. For instance, in the process described above, the probability that $2N_A$ particles are being in the left side or right side of a container is $\left(\frac{1}{2}\right)^{2N_A}$ as is clear in relation (6). This situation corresponds to a less probable state (as compared to the state that particles are distributed uniformly in the entire space of a container), but with more information about the spatial distribution of particles. Therefore, in some circumstances for which the relation (7) is satisfied, the entropy change indicates the transition from a more probable state (the state in which $\lambda N_A$ identical particles of *A* are distributed in the entire space with probability one) to a less probable state which can be attained with a probability of $\left(\frac{1}{\lambda}\right)^{\lambda N_A}$. This is a phenomenon with statistical nature which does not happen spontaneously and it seems that is not compatible with the common interpretation of the second law. Because, in our proposed experiment, the system does not exchange energy with environment and is supposed to be isolated. Yet, the main and fundamental question is that whether the second law restricts the occurrence of such processes.

To answer this inquiry, we first review the Gibbs extension of the second law to the processes involving the material change. Here, the differential definition of entropy is generalized as:

$$dS = \frac{dq}{T} - \frac{1}{T}\sum_j \mu_j dN_j \tag{8}$$

For an isolated system, $dq = 0$, and according to the second law $dS \geq 0$. So, $\sum_j \mu_j dN_j \leq 0$, where the equality sign applies only when the system is in the material equilibrium. One can also reach the same result, beginning with the Clausius enequality $dS \geq \frac{dq}{T}$. So the inequality $\sum_j \mu_j dN_j \leq 0$ could be considered as a general condition for the material equilibrium in spontaneous processes. Now, according to relation (8), for an *IGM* process, $dS = -\frac{1}{T}\sum_j \mu_j dN_j$ where for an ideal gas with identical particles we have

$$\mu_j = -kT \ln \frac{z_j}{N_j}; \quad z_j = z, \forall j \tag{9}$$

and $z_j$ is the partition function of particle *j*. Now, since $\sum_j dN_j = 0$, one gets:

$$\sum_j \mu_j dN_j = kT \sum_j \ln N_j dN_j = kT \sum_j d \ln N_j! \tag{10}$$

In above relation, the last term clearly shows the indistinguishability condition considered in Boltzmann's statistics. If this condition was not considered, the result of relation (10) would be zero which would lead to Gibbs' paradox.

In the process $b$, $dN_R = -dN_L$, where, subscripts $L$ and $R$ denote Left and Right respectively. Therefore:

$$\Delta(\mu_L dN_L + \mu_R dN_R) = kT \int_{N_A}^{2N_A} \ln(\frac{N_L}{2N_A - N_L}) dN_L = 2kT \ln 2^{N_A} \qquad (11)$$

which will lead to the same entropy change as in relation (6). One can see that, in this process, $\sum_j \mu_j dN_j > 0$ and $dS < 0$ which according to the Gibbs approach is in contradiction with the second law of thermodynamic.

But regarding the relation (10), it is clear that the second term in relation (8) has a statistical character in its essence. Furthermore, in the definition of the thermodynamic entropy (i.e., $dS = \frac{dq_{rev}}{T}$), the system is presumed to be closed (i.e. without material change) [6, 9]. Hence, one can reason that in Gibbs' extension of the second law two points are ignored:

1) the thermodynamic entropy is basically defined for closed systems without material change; and

2) in Gibbs' approach, the second term (i.e. $\frac{1}{T}\sum_j \mu_j dN_j$) has an implicit statistical character. For example, in a mixing process, this term appears because of the indistinguishability assumption for identical particles which has no meaning in ordinary thermodynamics.

From now on, by the second law of thermodynamics we mean the principle which describes the entropy change as $\Delta S_{th} = \int \frac{dq_{rev}}{T}$ for closed systems. For isolated systems, this calls for the fact that the change of *thermodynamic* entropy (i.e., $\Delta S_{th}$) should be always positive (for irreversible processes) or zero (for reversible processes). We restrict the applicability of the second law for the *thermodynamic* entropy which is a conventional definition of entropy when there is no change of matter. We admit also that there are many processes in which the change of thermodynamic entropy can be interpreted as a witness of change of information. For example, when an ideal gas is expanded into the vacuum in an adiabatic process, the information about the location of particles is diminished. So, the concept of information is enfolded in thermodynamic entropy too. But, there is a *special* information-based feature (described by relation (10)) that the second law (and therefore the thermodynamic entropy) does not include it directly.

Now, let's us give primary role to the statistical approach of thermodynamics. In this way, we can define statistical entropy which its infinitesimal change is assumed to be:

$$dS_{st} = dS_{th} - \frac{1}{T}\sum_j \mu_j dN_j \qquad (12)$$

where, $S_{st}$ is the statistical entropy and $S_{th}$ is the thermodynamic entropy appeared in relation $dS_{th} = \frac{dq_{rev}}{T}$. Then, regarding the relation (12) and noting that $dS_{th} \geq \frac{dq}{T}$ for a closed system, one gets:

$$dS_{st} \geq \frac{dq}{T} - \frac{1}{T}\sum_j \mu_j dN_j \qquad (13)$$

For an isolated system, $dS_{st} \geq -\frac{1}{T}\sum_j \mu_j dN_j$ where the equality holds when the system is in mechanical and thermal equilibrium. Here, one can see that the second law, by itself, does not characterize the condition of material equilibrium. In an isolated system if $\sum_j \mu_j dN_j \leq 0$, then $dS_{st} \geq 0$. Otherwise, $dS_{st}$ may be also negative. This situation, *ipso facto*, is not in contrast to the second law. Yet, as an important complementary rule, we can accept that the condition $\sum_j \mu_j dN_j \leq 0$ for *the spontaneous processes* is still characterizing the condition of material equilibrium.

The second term in relation (12) has a *special* statistical character as is explicitly demonstrated in relation (10). Our emphasis on "special" here means that without this term, the *indistinguishability* condition considered in Boltzmann's statistics would be meaningless. This is the characteristic feature of this term only. When the first term is zero (i.e., for an isolated system undergoing a reversible change of state), the change of the second term is responsible for the change of entropy. Here is the point that a distinctive notion of information is introduced in the definition of entropy, when the thermodynamic entropy is constant and its change is zero. So one can discriminate the notion of statistical entropy (with a special informational character for isolated systems) from the thermodynamic entropy described by the second law ($dS_{th} = \frac{dq_{rev}}{T}$) in relation (12).

Our approach, on the other hand, shows that on the basis of a statistical formulation of thermodynamic, the second law does not restrict the amount of information which one can obtain about a particular situation of the system (accompanied by a change of the thermodynamic state), whenever the process is conducted by a conscious observer without performing work. Therefore, this is a special information-based feature that the second law does not necessarily encompass it. Consequently, the final state of an isolated system could be controlled in such manner that eventually any probable state would be achieved. This change, of course, does not happen spontaneously, but can be occurred *in principle* without any exchange of the energy with environment. This could be regarded as a new aspect of entropy.

In effect, the system might have an information-based interaction with surrounding without heat-work interaction. When the information of an observer about the position of particles is increased (corresponding to a state of lower probability), $\Delta S_{st} < 0$, and when his or her information is decreased (corresponding to a state of higher probability), $\Delta S_{st} > 0$. The possibility of such events, in turn, is determined by the inequality $dS_{st} \geq -\frac{1}{T} \sum_j \mu_j dN_j$ which is an appropriate generalization of the second law.

## 5  Maxwell's Demon

Maxwell's Demon perhaps was the first idea for finding out how a relationship between entropy and information could be established [3]. In a paper published in 1929, Szilard proposed a principle that "the act of measurement which determines the position of molecules causes an increase in entropy, so that it compensates the entropy decrease which is due to the act of the Demon" [10]. On the other hand, to exorcise Maxwell's Demon, Brillouin mentioned the idea that gaining information (bound information) about a physical system decreases its physical entropy. He defined 'negentropy' ≡ N = −(entropy), so that $\Delta N = -\Delta S$. Then, he applied his negentropy principle of information to an isolated system and concluded that $\Delta(Entropy) - \Delta(Information) \geq 0$.

Szilard's insight was expanded upon in 1982 by Charles H. Bennett after that Rolf Landauer introduced the concept of the "logical irreversibility" in connection with information discarding processes in computers. Logical irreversibility means that starting from a given state one cannot get to another state without using further information (including, e.g., the computer program and some initial data). So, to determine what side of the gate a molecule must be on, the demon must store information about the state of the molecule. Eventually, the demon will run out of information storage space and must begin to erase the information that has been previously gathered. Erasing information, however, is assumed to be a thermodynamically irreversible process that increases the entropy of a system. (Erasing one bit of information produces $kT \ln 2$ dissipated heat) [12, 13]. The arguments of Brillouin and Landauer-Bennett about the relationship between the entropy and information have been recently criticized [14, 15].

The important point here is that by clarifying the concept of entropy and its relationship with information (as described above), the entropy can be decreased in a non-spontaneous process where there is no exchange of energy with environment. But the decrease in entropy does not violate the second law, as far as an intervening agent is involved. The validity of the second law neither necessitates the compensation of information to the environment (according to what Szilard claimed), nor (considering what Landauer-Bennett introduced) is inevitably related to the concept of irreversibility of the information erasure. In an exorcism approach, one is not obliged to presume the logical irreversibility assumption of the information erasure according to Landauer's principle, since there might be other possibilities like the one we explored in this paper. In effect, we are not going to disprove the Landauer principle but rather to present a novel and different view of the relationship between thermodynamic entropy and information.

We emphasized that the thermodynamic entropy does not generally include the processes in which there is material change. Therefore, for an isolated system,

$$\Delta S_{st} \geq -\Delta(\frac{1}{T}\sum_{j}\mu_j dN_j) \qquad (14)$$

In this situation, entropy need not be compensated, since there is no constraint on its value except for what the relation (14) demands.

In some other efforts, much work has been done on the relation of entropy and information in recent years. Yet, most of the authors think over the thermodynamic systems either under a quantum regime [16, 17, 18] or in the context of a quantum entanglement with a bath [19, 20, 21, 22]. A dominant presumption in all these works is that during the change of state of the system, there is no material change. What we are considering here, however, is a *macroscopic* system which is not in material equilibrium. Hence, we are surveying the relationship between entropy and information in a pure thermodynamic context.

To sum up, when there is material change, from an information-based aspect, the overall entropy of the system and environment is not constrained. This is in contrast to the situation where there is no material change. Hence, the Demon can change the state of the system to a state of lower entropy (corresponding to a state which its occurrence is less probable) and in this way his information increases.

We should, however, notice that here, for having an information-based interaction with environment without any exchange of energy, the existence of material change along with the assumption of indistinguishability of particles are two necessary conditions.

# 6 Conclusion

What is emphasized in this paper, at first, is to show a "possibility"; the possibility of the existence of some events in which the entropy of system changes from higher to lower values without any energy interaction with environment. In such non-spontaneous processes, in principle, entropy may decreases. But it doesn't mean the violation of the second law, because the thermodynamic entropy is basically not defined for processes in which there is material change. Instead, this result itself is deduced from an appropriate description of the second law in material processes. From a statistical point of view, the entropy could be defined in a more general form, so that only for the systems which are not involved in material processes, the two definitions (according to Relation (12)) coincide. That is, in some processes, a system may have information-based interactions with environment (containing conscious observer) without any exchange of energy. This is an important facet of entropy that should be discriminated from an energy feature. Hence, the increase or decrease of information of an intervening observer (contrary to what is generally supposed) is not restricted by the second law, although it has relationship with it.

Figure Captions:

Fig. 1. The process of dissociation of two ideal gases A and B to two equivalent parts with the same temperature and volume.

Fig. 2. The partitioning of an ideal gas A to two equivalent parts with the same temperature and volume.

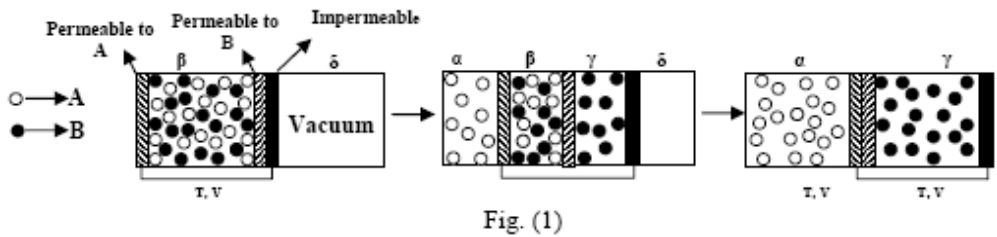

Fig. (1)

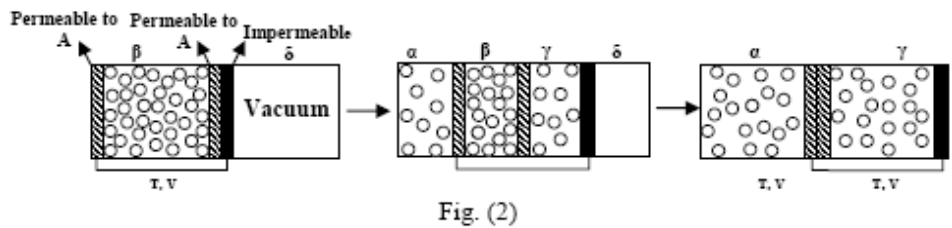

Fig. (2)